# Black hole as fireplace: limited communications across the horizon

Liangsuo Shu[1,2], Kaifeng Cui[3,4], Xiaokang Liu[2], Zhichun Liu[2], Wei Liu[2]

[1]School of Physics, Huazhong University of Science & Technology. Wuhan, China.
[2]School of Energy and Power Engineering, Huazhong University of Science & Technology. Wuhan, China.
[3]Key Laboratory of Atom Frequency Standards, Wuhan Institute of Physics and Mathematics, Chinese Academy of Sciences, Wuhan, China
[4]University of Chinese Academy of Sciences, Beijing, China

**Abstract**

An insightful viewpoint was proposed by Susskind about AMPS firewall: the region behind the firewall does not exist and the firewall is an extension of the singularity. In this work, we provided a possible picture of this idea by combining Newman's complex metric and Dvali-Gomez BEC black holes, which are Bose-Einstein condensates of $N$ gravitons. The inner space behind the horizon is a realized imaginary space encrusted by the real space outside the horizon. In this way, the singularity extents to the horizon to make a firewall for the infalling observer. Some gravitons escape during the fluctuation of the BEC black hole, resulting in a micro-transparent horizon which makes the firewall exposes slightly to an observer outside the horizon. This picture allows limited communications across the horizon.

L. S. : liangsuo_shu@hust.edu.cn    K. C. : cuikaifeng@wipm.ac.cn
X. L. : xk_liu@hust.edu.cn          Z. L. : zcliu@hust.edu.cn
W. L.: w_liu@hust.edu.cn

# 1. Introduction

The black hole information paradox shows the conflict between quantum mechanics and general relativity. Solving this paradox is likely to resolve the conflict and provide road signs to quantum gravity. Therefore, it has received lasting attentions [1–11]. Among them, AMPS firewall [5] has attracted much attention recently.

In the discussion about AMPS firewall [5], Susskind put forward an insightful viewpoint: the region behind the firewall does not exist and the firewall is an extension of the singularity [6]. Fuzzball [4] based on string also suggest that there is no singularity at the center of a black hole but a bag of string. This hypothesis also supported by a quantum pictures of black holes provided by Dvali and Gomez recently: black holes can be viewed as Bose-Einstein condensate (BEC) of $N=M^2$ soft constituent gravitons at the critical point of a quantum phase transition [12–16]. The gravitons in BEC are indistinguishable. Logically, as a composed of these indistinguishable gravitons, the space within the horizon should be the same point.

In a related work [17], we found that both black holes and elementary particles can be described by complex Kerr-Newman metric [18,19]. In fact, they are two special cases of the complex black holes and can convert to each other through a phase transition: after a phase transition point at Planck energy, the particle's imaginary radii are realized and converted into a black hole. In this work, by analyzing the gravitons of a Dvali-Gomez BEC black hole from the view of complex metric, the singularity of a black hole was found to be a polymer of the ring singularities of the gravitons it owns. In this picture, the singularity extents to the horizon to make a firewall for the infalling observer. However, the communication across the horizon at limited rate can happen when a part of its gravitons escape from the horizon temporarily during fluctuations. In other word, the black hole firewall is only slight-naked rather than the dangerous naked one in [20].

## 2. Inside Firewall as an extension of the singularity

### 2.1. Gravitons in realized imaginary space

In this section, we will show how Susskind's insightful viewpoint about firewall is achieved in Dvali-Gomez BEC black hole using complex Kerr-Newman metric [18,19].

From [17], we know the complex radius of a complex Kerr-Newman black hole can be described as

$$r = m \pm i\sqrt{a^2 + Q^2 - m^2} \tag{1}$$

where $m$ is its mass and energy, $a$ is its angular momentum per unit mass, and $Q$ is its charge, $c = \hbar = G = k_B = 1$ is used in this work. The real radius of the complex horizon ($r_R$) is,

$$r_R = m \tag{2}$$

while the imaginary radii of the complex horizon ($r_I$) are,

$$r_I = \pm i\sqrt{a^2 + Q^2 - m^2} \tag{3}$$

In low-energy, $a_m$ of an elementary particle is much larger than its $Q_m$ and $m$, $r_I$ is approximately equal to $ia_m$.

$$r_I = i\sqrt{a_m^2 + Q_m^2 - m^2} \approx ia_m \tag{4}$$

As the particle energy, $m$, increases, its imaginary radii decrease. After the point of phase transition, $r_I=0$, the imaginary radii are realized and the particle transform into a real Kerr-Newman black hole. Therefore, the time-like space between its two horizons is a realized imaginary space encrusted by the real space outside the horizon: $r_R$ of a real Kerr-Newman black hole describes its origin, appearing as a 2-D spherical surface in 3-D real space, while its realized $r_I$ determines the boundary of the realized imaginary space, appearing as its two horizons.

According to Dvali and Gomez's work [12], a graviton in the BEC Schwarzschild black hole has mass of

$$m = M/N = 1/M \tag{5}$$

Its spin, $L_m$, is 2 and $Q$ is 0, therefore

$$r_I = ia_m = 2Mi \qquad (6)$$

The module of the graviton's $r_I$, $2M$, is exactly the radius of the Schwarzschild black hole. The graviton's $r_R$,

$$r_R = m = 1/M \qquad (7)$$

is found to be the Compton wavelength of the black hole when it is treated as an elementary particle [21–33]. The Compton wavelength of a particle expresses a fundamental limitation on measuring its position [34].

In this picture, all the points of the imaginary space in the horizon share the same real coordinate. In other word, from the view of an observer in real space, there is in fact no difference between the center of a black hole and one point at its horizon. The singularity of a black hole can be regarded as a big polymer of the ring singularities of its gravitons. Therefore, the information of a Black hole can be thought of as being stored at its horizon as well as at its center, which is consistent with the holographic principle [35,36]. The singularity of the black hole extends from the center to the horizon. In this way, the firewall inside the horizon for an infalling observer is nothing but an extension of the singularity. In fact, the gravitons in BEC black hole can also be regarded as string with radius of 2M because their real radii are much smaller than the imaginary radii. This implies that fuzzball [4] and firewall [5] should have "genetic relationship".

## 2.2. Gravitons as ash of infalling observer

Once an infalling observer crosses the horizon, it will hit part of the polymer singularity: some ring singularities of its gravitons. If the infalling observer carrying no net charge and angular momentum, the ashes of it is a certain number of graviton.

$$dN = dM^2 = 2MdM \qquad (8)$$

where $dN$ is the increased number of the gravitons when the BEC black hole absorbs an infalling object with energy of $dM$. The total energy of the new graviton members is

$$mdN = \frac{1}{M + dM} 2MdM \tag{9}$$

when $dM$ is much smaller than $M$,

$$mdN \approx 2dM \tag{10}$$

Therefore, one half of the total energy of the new graviton members comes from the infalling object, while another half from the contribution of the original graviton members.

From the works of Bekenstein [34,37] and Hawking [38,39], we know the information stored in a Schwarzschild black hole with mass of $M$ for an outer observer is

$$S = 4\pi M^2 \tag{11}$$

Since there are $N$ indistinguishable gravitons in the BEC black hole, the information carried by every graviton is

$$S_m = S/N = 4\pi \tag{12}$$

The information carried by every graviton are not independent because these gravitons of BEC black hole are indistinguishable. Therefore, the BEC black hole should be a quantum entanglement system. The entanglement in BEC have been discussed in [40–43].

## 3. Outside Firewall as an appendage of the inside firewall

### 3.1. Temporarily escaping gravitons during fluctuation

The gravitons and their ring singularities should be hidden well by the horizon in an equilibrium black hole. Otherwise, the black hole will destroy the causality of its surrounding space and become an un-physical existence.

However, part of the gravitons may escape from the shade of the horizon through quantum tunneling [3] or quantum depletion of graviton condensate [15]. Regardless of the dynamic mechanism, in terms of thermodynamics, this process is an entropy reducing fluctuation of the black hole. The fluctuation of a system cannot be avoided because of the statistical nature of the second law of thermodynamics. Assuming $dN$ gravitons escape from the horizon temporarily during the fluctuation, from equations (8) and (11), the energy loss of the black hole will be

$$dM = \frac{dN}{2M} \qquad (13)$$

and the information loss of the black hole will be

$$dS = 8\pi M dM = 4\pi dN = S_m dN \qquad (14)$$

According to the fluctuation theorem [44], the relative probability of the fluctuation

$$P_-/P_+ = e^{-dS} \qquad (15)$$

where $P_-$ is the probability of the fluctuation from $S$ to $S-dS$, while $P_+$ is the probability of its reverse process. The sum of $P_-$ and $P_+$ is 1, therefore

$$P_- = \frac{1}{1+e^{dS}} \qquad (16)$$

If $dS$ is much larger than 1, $P_-$ will be approximately equal to

$$P_- \approx e^{-dS} \qquad (17)$$

The result of above thermodynamic analysis, in agreement with [3], implies that the temporarily escaped gravitons have a non-thermal distribution. Equation (17) also can be obtained from quantum field theory via Fermi's Golden Rule [45]. Therefore, the unitarity stands up during the fluctuation if equation (17) holds.

After crossing the horizon, the temporarily escaping gravitons during the fluctuation lose half of their energy, which means there is a potential barrier at the horizon. Since the remaining graviton shared the energy left behind, the potential barrier at the horizon should be a result of the interaction between the escaping gravitons and remaining ones. After the energy recovery, the increase of $m$ of the remaining graviton is

$$dm = d\frac{1}{M} = -\frac{dM}{M^2} \qquad (18)$$

## 3.2. Entanglement swapping creates a firewall and send information

The temporarily escaping gravitons are entangled with the black hole before the quantum entanglement between them is destroyed. If these gravitons do not interact with other particles during the fluctuation, they will return to the black hole because $P_+$ is much larger than $P_-$. If Hawking radiation is to take away the information carried by the temporarily escaping gravitons, it must interact with them before they return to the horizon. Inspired by quantum information [46] and information thermodynamics [47], we propose a possible solution which respects the basic principles of quantum mechanics.

From Hawking's works [38,39], we know the vacuum fluctuation near the horizon, causing particle–antiparticle pairs, will result in Hawking radiation. The particle and its antiparticle of a Hawking pair are entangled with each other. If the antiparticle with negative energy and one graviton temporarily escaped from the horizon are made complete Bell measurement, the particle will copy the quantum state of the graviton. This is in fact a progress of entanglement swapping [46], In this way, the Hawking radiation will take the information of the black hole away and lose its black body spectrum. After the entanglement swapping, the particle of Hawking radiation is entangled with the black hole. Throughout the process, both the no cloning theorem and the monogamy of entanglement stand up.

In the above hypothetical entanglement swapping, what acts as the observer doing the complete Bell measurement? This measurement is essentially an information processing process. According to Landauer principle [47], which was found to stand up to quantum test [48,49], it has to consume energy. Therefore, the heat released by the complete Bell measurement can create an outside firewall near the horizon for an out-escaping observer. Treating the Bell measurement in this firewall as a black box process, we can find that the role played by the firewall is to destroy the old quantum entanglement between the two particles of the Hawking pair and create new quantum entanglement between the particle and the black hole, which is exactly the function of the AMPS firewall [5]. As a result of the micro-transparent horizon caused by the fluctuation, the outside firewall can be regarded as an appendage of the inside firewall.

## 4. Conclusion and discussion

The starting point of this work is an insightful viewpoint proposed by Susskind about AMPS firewall: the region behind the firewall does not exist and the firewall is an extension of the singularity [6].

In the first part, by combining Dvali-Gomez BEC black hole [12–16] and Newman's complex Kerr-Newman metric [18,19], we provided a possible picture of Susskind's idea: the space behind the horizon is a realized imaginary space, which is encrusted by the real space outside the horizon and share the same real coordinate. Therefore, for a real observer outside the horizon, there is no difference between the center and the horizon of a Schwarzschild Black Hole. In this way, the singularity extents to the horizon to make an inside firewall. Once an infalling object cross the horizon, it will be burned to ash: gravitons of the BEC black hole.

In the second part, we show a possible mechanism how Hawking radiation takes away the information of black hole through entanglement swapping. The inside firewall is hidden well by the horizon. However, during a fluctuation of the BEC black hole, some gravitons can escape temporarily. In this way, for an outside observer, the horizon become micro-transparent and the inside firewall exposes slightly. Then the particles of Hawking radiation interact with the escaping gravitons and get their information through entanglement swapping. According to Landauer principle [47], the entanglement swapping as an information processing process has to consume energy. The released heat creates an outside firewall, which destroys the old quantum entanglement between the two particles of the Hawking pair and create new quantum entanglement between the particle of Hawking radiation and the black hole.

In summary, the Hawking radiation process can be summarized as the following three main processes:
1. The singularity of a black hole extents to horizon and makes an inside firewall. Once an infalling object crosses the horizon, it will be burned to ash: gravitons of the black hole.

2. Some gravitons escape during the fluctuation of the BEC black hole, resulting in a micro-transparent horizon which makes the inside firewall exposes slightly for an observer outside the horizon.

3. The particles of Hawking radiation interact with the escaping gravitons and get their information through entanglement swapping. The heat released by this information processing process creates an outside firewall.

This mechanism not only well hides the inside firewall, an extension of the singularity, but also allows the information of the black hole to be transmitted at a limited rate. In this way, the black hole is like a gravitational fireplace with a wall of the horizon. In the fireplace, it is the danger fire of gravitons and their ring singularities. Once an infalling observer crosses the wall, it will become the fuel of the fireplace and burnt to ash of gravitons. However, fluctuation makes the wall of the fireplace micro-transparent, resulting in that an outside observer can both feel warm and see the flame, from the color of which the observer may restore some information of the fuel supplying for the fireplace.

## Acknowledgements

This works is supported by the National Science Foundation of China (No. 51736004 and No.51776079).